\documentclass[aoas,nameyear,dvips]{arximspdf}
\usepackage{stfloats}
\usepackage[dvips,all]{xy}
\usepackage{xypic}
\usepackage{graphicx}

\doi{10.1214/10-AOAS372}
\volume{5}
\issue{4}
\pubyear{2011}
\firstpage{2300}
\lastpage{2325}

\makeatletter
\fnbelowfloat
\makeatother

\begin{document}
\begin{frontmatter}

\title{Simultaneous analysis of a sequence of paired ecological
tables: A comparison of several methods}
\runtitle{Analysis of a Sequence of Paired Ecological Tables}

\begin{aug}
\author{\fnms{Jean} \snm{Thioulouse}\corref{}\ead[label=e1]{jean.thioulouse@univ-lyon1.fr}}
\address{Universit\'{e} de Lyon\\
 F-69000, Lyon and Universit\'{e} Lyon 1 CNRS, UMR5558\\
Laboratoire de Biom\'{e}trie et Biologie Evolutive F-69622\\
Villeurbanne Cedex\\
France\\
\printead{e1}}
\affiliation{University of Lyon and CNRS}
\runauthor{J. Thioulouse}
\end{aug}

\received{\smonth{10} \syear{2009}}
\revised{\smonth{6} \syear{2010}}

%
\begin{abstract}
A pair of ecological tables is made of one table containing
environmental variables (in columns) and another table
containing species data (in columns). The rows of these two tables are
identical and correspond to the sites where
environmental variables and species data have been measured. Such data
are used to analyze the relationships between
species and their environment. If sampling is repeated over time for
both tables, one obtains a sequence of pairs of
ecological tables. Analyzing this type of data is a way to assess
changes in species-environment relationships,
which can be important for conservation Ecology or for global change
studies. We present a new data analysis method
adapted to the study of this type of data, and we compare it with two
other methods on the same data set. All three methods are implemented
in the ade4 package for the R environment.
\end{abstract}

%

\end{frontmatter}

\section{Introduction}\label{intro}

Ecological data analysis has been very productive in the second part of
the 20{th} century. Many original multivariate data analysis methods
have been developed, particularly those designed to tackle the
fundamental issue of Ecology: the description of the relationships
between species and their environment.

These methods study the relationships between species and their
environment through two data tables, called a ``pair of ecological
tables.'' The first of these two tables contains environmental
variables (in columns) recorded in a set of sampling sites (rows).
These variables are usually quantitative, physico-chemical properties,
for example, and they can also be categorical.

The second table of the pair is the species table, containing species
data recorded at the same sampling sites. This can be the number of
organisms, their presence/absence, or an abundance level. The rows in
this\vadjust{\goodbreak} table correspond to the sampling sites, and its columns correspond
to the species. \citet{doledec94} present a short review of linear
ordination methods for studying species--environment relationships, and
the paper by \citet{dray03} features a comparison of the advantages
and disadvantages of recent methods.

Ecologists are also interested in the \emph{changes} in the
relationships between species and environment [\citet{guisan00}].
Indeed, variations of these relationships can be important, for
example, from the point of view of species conservation or for global
change studies. When sampling is repeated in time (or in space), one
gets a sequence of tables, also called a~$k$-table. When a pair of
ecological tables is repeated, the result is a pair of $k$-tables, or two
data cubes. One sequence of tables makes one data cube, and a sequence
of pairs of tables makes a pair of data cubes (a species data cube plus
an environmental data cube). Analyzing the relationships between the
two cubes can give useful insights into the evolution of
species--environment relationships.

From the point of view of statistical methods, two approaches can be
contrasted: the descriptive strategy and the predictive strategy. The
aim of the first one is an objective description of the data set and of
the relationships between its components. The second approach is
orientated toward the prediction of ``explained'' (or ``dependent'')
variables by ``explanatory'' (or ``independent'') ones. This
distinction implies an asymmetry of predictive methods and a symmetry
of descriptive ones. Indeed, descriptive methods do not differentiate
between ``explained'' and ``explanatory'' variables.

This difference has consequences on computational constraints:
predictive methods have a matrix inversion step that is not present in
descriptive methods. This matrix inversion step has negative
consequences on the data sets that can be analyzed: it means that
``explanatory'' variables must be independent (in the statistical
sense), {that is}, that they must be linearly independent,
because the rank of their correlation matrix must not be less than its
dimension. This also implies that the number of cases (samples) must
not be less than the number of explanatory variables.

This constraint is particularly important since the advent of
bioinformatics, with the huge data sets provided by high throughput
molecular biology methods like DNA microarrays and DNA fingerprints.
These methods can produce extremely large data tables with very low
information density. There are potentially thousands of variables,
corresponding, for example, to electrophoresis bands or to DNA sequence tags.

\begin{figure}[t]

\includegraphics{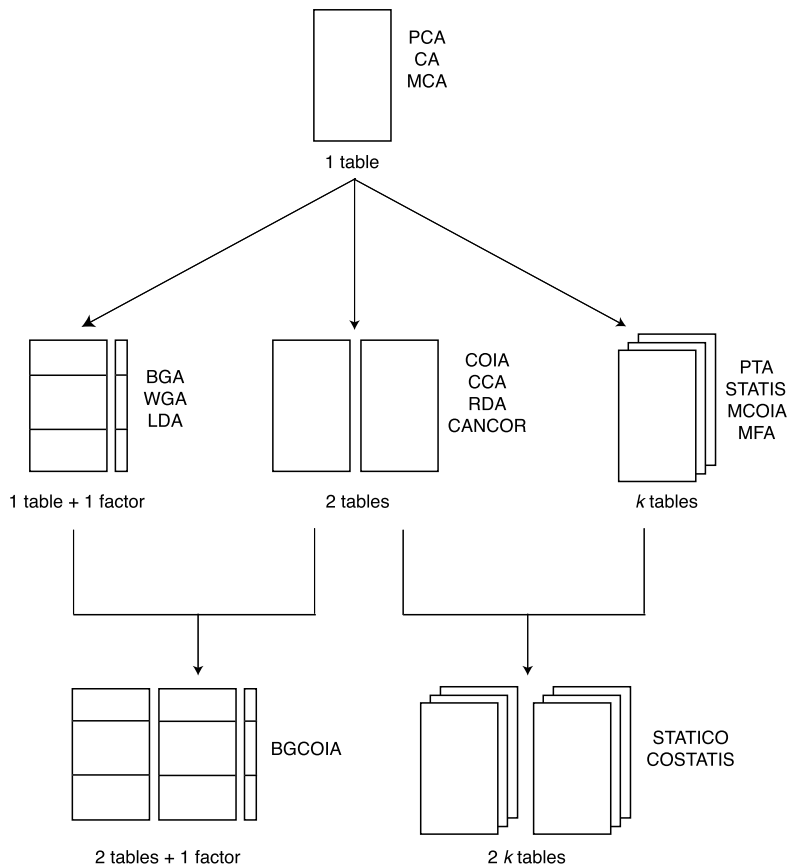}

\caption{Diagram describing data structures in various experimental
conditions: one table, one table with groups of rows, two tables, one
k-table, two tables with groups of rows, and two k-tables. Data
analysis methods corresponding to these situations are given on the
right of each data structure. Abbreviations are given in the text (see
\protect\hyperref[intro]{Introduction}).}\label{fig1}
\end{figure}

Conversely, descriptive methods can be used without constraint on the
ratio between the number of samples and the number of variables. The
main body of this paper is restricted to this approach. Figure~\ref{fig1} shows
a diagram describing the methods that we are going to present, with the
corresponding data structures. Abbreviations are given in the rest of
this introduction.

In the context of ecological data analysis, the distinction between
predictive and descriptive methods is particularly important in two
cases: when samples (the rows of data tables) belong to groups, and in
the case of a pair of ecological tables.

When \textit{groups of samples} are involved, Linear Discriminant
Analysis (LDA) [\citet{Venables02}] is a classical example of a
predictive approach. Between-Group Analysis (BGA)
[\citet{doledec87}; \citet{culhane02}] is a descriptive method analogous to LDA, but it
can be used even when the number of samples is less than the number of
variables. It can be seen simply as the PCA of the table of group
means. Within-Group Analysis (WGA) [\citet{doledec87}] is the reverse of
Between-Group Analysis: it is the PCA of the residuals between initial
data and group means. It removes the effect of the grouping variable
and analyzes the remaining variability.

The standard for the analysis of \textit{one pair} of ecological
tables is Canonical Correspondence Analysis (CCA) [\citet{terbraak86},
not to be confused with Canonical Correlation Analysis (CANCOR)].
Canonical Correspondence Analysis can be seen as a particular form of
Correspondence Analysis (CA) [\citet{benzecri73}; \citet{hill73}] or of
Multiple Correspondence Analysis (MCA) [\citet{tenenhaus85}], where
sample scores are constrained to be a linear combination of
environmental variables. It belongs to the predictive approach, and
involves a regression step (including a matrix inversion). It is
therefore restricted to the case where explanatory variables (usually
environmental parameters) are linearly independent and not too many.
Redundancy Analysis (RDA) [\citet{legendre98}] is similar to Canonical
Correspondence Analysis, but it is a constrained PCA instead of a
constrained Correspondence Analysis.

On the other hand, Co-Inertia Analysis (COIA) [\citet{doledec94};
\citet{dray03}] belongs to the descriptive approach. It a simple and robust
alternative to Canonical Correspondence Analysis when the number of
samples is low compared to the number of explanatory (environmental)
variables. Co-Inertia Analysis can be seen as the PCA of the table of
cross-covariances between the variables of the two tables. Other
advantages are detailed in \citet{dray03}.

$K$-table analysis methods are used to analyze \textit{series} of
tables. They belong to three families: STATIS [\citet{lavit94};
\citet{escoufier73}], Multiple Factor Analysis (MFA) [\citet{escofier94}], and
Multiple Co-Inertia Analysis (MCOIA) [\citet{chessel96}]. Partial
Triadic Analysis (PTA) [\citet{thioulouse87}] is one of the simplest
analyses of the STATIS family, and it can be seen as the PCA of a
series of PCAs.

Multivariate analysis methods for \textit{pairs of data cubes} are not
widespread. Two of them are based on co-inertia: the first one is
called Between-Group Co-Inertia Analysis (BGCOIA) [\citet{franquet95}],
and the second one is the STATICO method [\citet{simier99}; \citet{thioulouse04}].
In this paper we present a new method called
COSTATIS, and we compare it with BGCOIA and STATICO. The name STATICO
means ``STATIS and CO-inertia,'' while COSTATIS means ``CO-inertia and
STATIS.'' STATIS is a French abbreviation for ``Structuration des
TAbleaux \`a Trois Indices de la Statistique'' (organization of three
way tables in Statistics).

The comparison of the three methods is done on the same data set as the
one used by \citet{franquet95} and by \citet{thioulouse04}. BGCOIA and
STATICO have already been presented in biological journals, so we
briefly present their methodological bases. The COSTATIS method has
never been presented before, so we explain it here. We compare the
results of the three methods from a rather practical point of view, on
their respective graphical outputs for the same data set, and on their
global properties.

Functions for the R software [\citet{rcoreteam}] to perform
computations and graphical displays for the three methods are available
in the ade4 package [\url{http://pbil.univ-lyon1.fr/ade4/} see \citet{chessel04};
 \citet{dray07}]. All the computations and graphical
displays can be redone interactively online, thanks to this
reproducibility page: \url{http://pbil.univ-lyon1.fr/SAOASOPET/}.

\section{Example data set and basic analyses}\label{sec2}

The three data cube coupling methods presented here are based on three
different basic analyses: Between-Group Analysis, Co-Inertia Analysis
and Partial Triadic Analysis.\footnote{BGCOIA is based on Between-Group
Analysis plus Co-Inertia Analysis, STATICO is based on Co-Inertia
Analysis plus Partial Triadic Analysis and COSTATIS is based on
Partial Triadic Analysis plus Co-Inertia Analysis.} In this section we
give a short summary of these basic analyses in the framework of the
duality diagram. We start the section by presenting a description of
the example data set.

\begin{figure}

\includegraphics{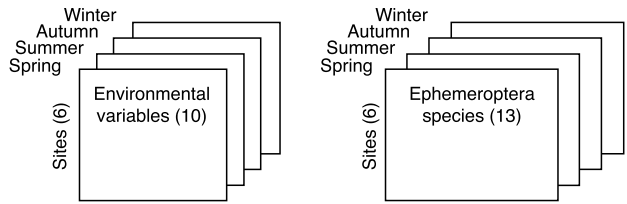}

\caption{The example data set consists of two data cubes. The first
one contains 10 environmental variables that have been measured four
times (in Spring, Summer, Autumn and Winter) along six sampling sites.
The second one contains the numbers of Ephemeroptera belonging to 13
species, collected in the same conditions.}\label{fig2}
\end{figure}

\subsection{Example data set}

We present the results of the three data cube coupling methods using
graphical displays obtained on the same data set. This data set is the
one used by \citet{franquet95} and by \citet{thioulouse04}. Numerical
data are printed in both papers, and they are also available in the R
package ``ade4,'' in the ``meau'' data set. A picture of these data as
two data cubes is given in Figure~\ref{fig2}. They are arranged in two tables:
one table with 24 rows and 10 columns, containing the environmental
variables, and one table with 24 rows and 13 columns, containing the
species data.

The rows of both tables correspond to 6 sampling sites ordered
upstream-downstream along a small stream, the M\'eaudret (South-East of
France, in the Vercors massif). Site 6 is a control, located on another
stream, the Bourne, into which the M\'eaudret flows. These 6 sites are
sampled 4 times, in Spring, Summer, Autumn and Winter. The 10
environmental variables are physico-chemical measures: water
temperature, flow, pH, conductivity, oxygen, biological oxygen demand
(BOD5), oxidability, ammonium, nitrates and phosphates. Most of these
variables are related to water pollution. Indeed, there is a large
summer mountain resort (Autrans) located between sites 1 and 2, and its
influence is predominant.

The 13 columns of the species data table correspond to 13 Ephemeroptera
species (mayflies), which are known to be highly sensitive to water
pollution. These species are as follows:

Eda $=$ \textit{Ephemera danica}, Bsp $=$ \textit{Baetis sp.}, Brh $=$
\textit{Baetis rhodani}, Bni $=$ \textit{Baetis niger}, Bpu $=$ \textit
{Baetis pumilus}, Cen $=$ \textit{Centroptilum}, Ecd $=$ \textit
{Ecdyonurus}, Rhi $=$ \textit{Rhithrogena}, Hla $=$ \textit{Habrophlebia
lauta}, Hab $=$ \textit{Habroleptoides modesta}, Par $=$ \textit
{Paraleptophlebia}, Cae $=$ \textit{Caenis}, Eig $=$ \textit{Ephemerella ignita}.

The goal of the analysis of this data set is to enlighten the
relationships between Ephemeroptera species distribution and the
quality of water. More precisely, data analysis methods should help
discover how these relationships vary in the space--time experimental
setup ({i.e.,} according to seasons during one year and along
the stream).

\subsection{PCA duality diagram}

The duality diagram [\citet{escoufier87}; Hol\-mes (\citeyear{holmes06})] will be
used in this paper to present several methods, so we explain it here
first for a simple Principal Component Analysis (PCA). Let $\mathbf
{X}=[x_{ij}]_{(n,p)}$ be the data table (environmental variables table,
for example), with $n$ rows (sampling sites) and $p$ columns
(variables). $\mathbf{X}^T$ is the transpose of $\mathbf{X}$. Let
$\mathbf{D}_n$ be the diagonal matrix ($n \times n$) of site weights:
$\mathbf{D}_n=\operatorname{diag}(w_1, \ldots, w_n)$, and let $\mathbf{D}_p$ be a
metric on~$\mathbb{R}^p$. The duality diagram of the general analysis
of the data table is as follows:

\begin{displaymath}
\begin{array}{c}
\xymatrix@L=8pt{
\fbox{$\mathbb{R}^p$} \ar[r]^{\mathbf{D}_p} & \fbox{$\mathbb{R}^{{p}^*}$} \ar
[d]^{\mathbf{X}\,.} \\
\fbox{$\mathbb{R}^{{n}^*}$} \ar[u]^{\mathbf{X}^T} & \fbox{$\mathbb{R}^n$} \ar
[l]^{\mathbf{D}_n} \\}
\end{array}
\end{displaymath}

This is called a ``duality diagram'' because ${\mathbb{R}^{{p}^*}}$
and ${\mathbb{R}^{{n}^*}}$ are the dual spaces of ${\mathbb{R}^p}$
and ${\mathbb{R}^n}$, and because the dual operators $\mathbf
{X}^T\mathbf{D}_n\mathbf{X}\mathbf{D}_p$ and $\mathbf{X}\mathbf
{D}_p\mathbf{X}^T\mathbf{D}_n$ share the same spectrum. This diagram
is completely defined by the ``triplet notation'': ($\mathbf{X},
\mathbf{D}_p, \mathbf{D}_n$), and the total inertia of this
statistical triplet is
\[
I_\mathbf{X}=\operatorname{trace}(\mathbf{X}\mathbf{D}_p\mathbf{X}^T\mathbf{D}_n).
\]

The generalized PCA (gPCA) of this triplet corresponds to the spectral
decomposition of $\mathbf{X}^T\mathbf{D}_n\mathbf{X}\mathbf{D}_p$.
When $\mathbf{D}_n$ is the matrix of uniform row weights ($w_i=1/n$),
and $\mathbf{D}_p$ is the identity (Euclidean metric), then this
analysis is a simple PCA, and if the variables are centered, the
inertia is the sum of their variances.

This duality diagram can be seen as a picture of the underlying
mathematical objects used in the theoretical description of the
analysis. It has several functions, like making it easier to remember
the characteristics of particular methods (mnemonic), finding out the
matrices that are needed to perform computations (for example, doing
one way around the diagram gives the matrix from which eigenvalues and
eigenvectors should be extracted), or where particular objects (like
row scores and variable loadings) can be found and how to compute them.
\citet{draydufour07} give a detailed description of the use of the
duality diagram in multivariate ecological data analysis and in the
ade4 package for the R environment.

\subsection{Between-Group Analysis}

Between-Group Analyses [Dol{\'e}dec and\break Chessel (\citeyear{doledec87}); \citet{culhane02}] are a
particular class of analyses, similar in their aim to linear
discriminant analysis, but comprising no matrix inversion step. They
consist in the summary (for example, by a PCA) of the table of group
means. In a second step, the rows of the initial table are projected in
this PCA to get row scores for all observations. The advantage is that
there is no constraint on the number of observations compared to the
number of variables, and no problem with numerous and/or correlated
variables, as it is the case in LDA. There are several types of
between-group analyses, corresponding to the initial analysis after
which Between-Group Analysis is computed. This can be, for example, a
PCA, a Correspondence Analysis, or a Multiple Correspondence Analysis.

In Between-Group Analysis, samples belong to $g$ groups, namely, $G_1,
\ldots,\allowbreak G_g$, with group counts $n_1, \ldots, n_g$, and $\sum
{n_k}=n$. Between-Group Analysis is the analysis of triplet $(\mathbf
{X}_B,\mathbf{D}_p,\mathbf{D}_{g})$, where $\mathbf{X}_B$ is the
$(g,p)$ matrix of group means:
\[
\mathbf{X}_B=[\bar{x}_j^k]_{(g,p)}.
\]
The term $\bar{x}_j^k=\frac{1}{n_k}\sum_{i\in{G_k}}{x_{ij}}$ is the mean
of variable $j$ in group $k$. In matrix notation, if $\mathbf{B}$ is
the matrix
of class indicators, $\mathbf{B}=[b_i^k]_{(n,g)}$, with $b_i^k=1$ if
$i\in{G_k}$ and
$b_i^k=0$ if $i\notin{G_k}$, then we have
\[
\mathbf{X}_B=\mathbf{D}_{g}\mathbf{B} ^T\mathbf{X}.
\]

Matrix $\mathbf{D}_{g}=\operatorname{Diag}(\frac{1}{n_k})$ is the diagonal matrix of
(possibly nonuniform) group weights, and $\mathbf{B}^T$ is the
transpose of $\mathbf{B}$. The corresponding duality diagram is the following:
\begin{displaymath}
\begin{array}{c}
\xymatrix@L=8pt{
\fbox{$\mathbb{R}^p$} \ar[r]^{\mathbf{D}_p} & \fbox{$\mathbb{R}^{{p}^*}$} \ar
[d]^{\mathbf{X}_B\,.} \\
\fbox{$\mathbb{R}^{{g}^*}$} \ar[u]^{\mathbf{X}_B^T} & \fbox{$\mathbb{R}^g$} \ar
[l]^{\mathbf{D}_{g}} \\}
\end{array}
\end{displaymath}

Between-Group Analysis is therefore the analysis of the table of group
means, leading to the diagonalization of matrix $\mathbf{X}_B^T\mathbf
{D}_{g}\mathbf{X}_B\mathbf{D}_p$. It's aim is to highlight the
differences between groups, and row scores maximize the between-group
variance. The statistical significance of these differences can be
tested by a permutation test, with a criterion equal to the
between/total variance ratio. Row scores of the initial data table can
be computed by projecting the rows of table $\mathbf{X}$ on the
principal component subspaces.

\subsection{Co-Inertia Analysis}

The first presentation of Co-Inertia Analysis dates back to \citet
{doledec94}, but almost ten years later, \citet{dray03} gave a more
detailed presentation and compared it with Canonical Correspondence
Analysis. Just as inertia is a sum of variances, co-inertia is a sum of
squared covariances, and Co-Inertia Analysis describes the co-structure
between two ecological data tables by summarizing as well as possible
the squared covariances between species and environment. Here is a
short description of this analysis.

Let $\mathbf{X}$ be the first table (environment variables table),
with $n$ rows (sampling sites) and $p$ columns (variables), and let
$\mathbf{Y}$ be the second table (species data), with the same $n$
rows and $q$ columns (species). $\mathbf{X}^T$ and $\mathbf{Y}^T$ are
the transpose of $\mathbf{X}$ and~$\mathbf{Y}$. Let $\mathbf{D}_n$
be the diagonal matrix ($n \times n$) of site weights: $\mathbf
{D}_n=\operatorname{diag}(w_1, \ldots, w_n)$, and let $\mathbf{D}_p$ and $\mathbf{D}_q$
be two metrics on $\mathbb{R}^p$ and $\mathbb{R}^q$ respectively.
Before doing the Co-Inertia Analysis, we need to analyze each table
separately. The duality diagrams of the separate analyses of the two
data tables are as follows:
\begin{displaymath}
\begin{array}{c}
\xymatrix@L=8pt{
\fbox{$\mathbb{R}^p$} \ar[r]^{\mathbf{D}_p} & \fbox{$\mathbb{R}^{{p}^*}$} \ar
[d]^{\mathbf{X}\,,} \\
\fbox{$\mathbb{R}^{{n}^*}$} \ar[u]^{\mathbf{X}^T} & \fbox{$\mathbb{R}^n$} \ar
[l]^{\mathbf{D}_n} \\}
\end{array}\hspace*{4pt}
\begin{array}{c}
\xymatrix@L=8pt{
\fbox{$\mathbb{R}^q$} \ar[r]^{\mathbf{D}_q} & \fbox{$\mathbb{R}^{{q}^*}$} \ar
[d]^{\mathbf{Y}\,.} \\
\fbox{$\mathbb{R}^{{n}^*}$} \ar[u]^{\mathbf{Y}^T} & \fbox{$\mathbb{R}^n$} \ar
[l]^{\mathbf{D}_n} \\}
\end{array}
\end{displaymath}

A generalized PCA of these triplets corresponds to the spectral
decomposition of $\mathbf{X}^T\mathbf{D}_n\mathbf{X}\mathbf{D}_p$
and $\mathbf{Y}^T\mathbf{D}_n\mathbf{Y}\mathbf{D}_q$. When $\mathbf
{D}_n$ is the\vadjust{\goodbreak} matrix of uniform row weights ($w_i=1/n$), and $\mathbf
{D}_p$ and $\mathbf{D}_q$ are identity (Euclidean metrics), then these
analyses are simple PCA.

Co-Inertia Analysis is defined by the duality diagram obtained by
merging these two separate diagrams. This will be possible when they
have the same spaces $\mathbb{R}^n$ and $\mathbb{R}^{n^*}$ in common,
which implies that the rows of the two tables must be identical. The
``coupled diagram'' of Co-Inertia Analysis is therefore
\begin{displaymath}
\begin{array}{c}
\xymatrix@L=8pt{
\fbox{$\mathbb{R}^p$} \ar[d]_{\mathbf{D}_p} & \fbox{$\mathbb{R}^{{n}^*}$} \ar
[l]_{\mathbf{X}^T} \ar[r]^{{\mathbf{Y}}^T} & \fbox{$\mathbb{R}^q$} \ar
[d]^{\mathbf{D}_q\,.} \\
\fbox{$\mathbb{R}^{{p}^*}$} \ar[r]_{\mathbf{X}} & \fbox{$\mathbb{R}^n$} \ar
[u]_{\mathbf{D}_n} & \fbox{$\mathbb{R}^{{q}^*}$} \ar[l]^{\mathbf{Y}} \\}
\end{array}
\end{displaymath}

Co-Inertia Analysis is the eigenanalysis of matrix $\mathbf
{X}^T\mathbf{D}_n\mathbf{Y}\mathbf{D}_q\mathbf{Y}^T\mathbf
{D}_n\mathbf{X}\mathbf{D}_p$ (starting in $\mathbb{R}^p$). This is
equivalent to the following ``crossed diagram'':
\begin{displaymath}
\begin{array}{c}
\xymatrix@L=8pt{
\fbox{$\mathbb{R}^p$} \ar[r]^{\mathbf{D}_p} & \fbox{$\mathbb{R}^{{p}^*}$} \ar
[d]^{\mathbf{Y}^T\mathbf{D}_n\mathbf{X}\,.} \\
\fbox{$\mathbb{R}^{{q}^*}$} \ar[u]^{\mathbf{X}^T\mathbf{D}_n\mathbf{Y}} &
\fbox{$\mathbb{R}^q$} \ar[l]^{\mathbf{D}_q} \\}
\end{array}
\end{displaymath}

This diagram highlights the fact that Co-Inertia Analysis is the
analysis of a cross product table, and its triplet notation is
($\mathbf{Y}^T\mathbf{D}_n\mathbf{X}, \mathbf{D}_p, \mathbf
{D}_q$). If the columns of both tables are centered, then the total
inertia of each table is simply a sum of variances: $I_\mathbf
{X}\!=\!\operatorname{trace}(\mathbf{X}\mathbf{D}_p\mathbf{X}^T\mathbf{D}_n)$ and
$I_\mathbf{Y}\!=\!\operatorname{trace}(\mathbf{Y}\mathbf{D}_q\mathbf{Y}^T\mathbf
{D}_n)$. And the co-inertia between $\mathbf{X}$ and $\mathbf{Y}$ is
in this case a sum of squared covariances:
\[
\mathit{CoI}_\mathbf{XY}=\operatorname{trace}(\mathbf{X}\mathbf{D}_p\mathbf{X}^T\mathbf
{D}_n\mathbf{Y}\mathbf{D}_q\mathbf{Y}^T\mathbf{D}_n).
\]

Co-Inertia Analysis maximizes the covariance between the row scores of
the two tables [\citet{dray03}]. Co-inertia is high when the values in
both tables are high simultaneously (or when they vary inversely) and
low when they vary independently or when they do not vary. This is
interesting from an ecological point of view: Co-Inertia Analysis will
show species that are abundant when some environmental variables are
particularly high (or low), and it will discard species that are not
influenced by these environmental variable. This is the meaning of the
co-structure between the two data tables.

The above ``coupled diagram'' shows the similarity of Co-Inertia
Analysis with Canonical Correlation Analysis (CANCOR). Indeed, the only
difference between the Co-Inertia Analysis and Canonical Correlation
Analysis duality diagram [\citet{caillez76}, p. 352; \citet{holmes06}]
comes from the metrics on $\mathbb{R}^p$ and $\mathbb{R}^q$:
\begin{displaymath}
\begin{array}{c}
\xymatrix@L=8pt{
\fbox{$\mathbb{R}^p$} \ar[d]_{\mathbf{V}^{-1}_{11}} & \fbox{$\mathbb{R}^{{n}^*}$}
\ar[l]_{\mathbf{X}^T} \ar[r]^{{\mathbf{Y}}^T} & \fbox{$\mathbb{R}^q$} \ar
[d]^{\mathbf{V}^{-1}_{22}\,.} \\
\fbox{$\mathbb{R}^{{p}^*}$} \ar[r]_{\mathbf{X}} & \fbox{$\mathbb{R}^n$} \ar
[u]_{\mathbf{D}_n} & \fbox{$\mathbb{R}^{{q}^*}$} \ar[l]^{\mathbf{Y}} \\}
\end{array}
\end{displaymath}

Canonical Correlation Analysis uses the Mahalanobis metric on $\mathbb
{R}^p$ and~$\mathbb{R}^q$, whose matrices are the inverse of the
covariance matrices $\mathbf{V}_{11}=\mathbf{X}^T\mathbf{D}_n\mathbf
{X}$ and $\mathbf{V}_{22}=\mathbf{Y}^T\mathbf{D}_n\mathbf{Y}$. This
leads to the Canonical Correlation Analysis triplet: ($\mathbf
{Y}^T\mathbf{D}_n\mathbf{X}, (\mathbf{X}^T\mathbf{D}_n\mathbf
{X})^{-1}, (\mathbf{Y}^T\mathbf{D}_n\mathbf{Y})^{-1}$). Canonical
Correlation Analysis row scores maximize their correlation, but this
can be achieved with very small variances. By maximizing the covariance
instead of the correlation, Co-Inertia Analysis ensures that the scores
do not have very small variances, and therefore have a good percentage
of explained variance in each space.

This diagram also clarifies the link between Co-Inertia Analysis and
instrumental variable methods like Principal Component Analysis with
respect to Instrumental Variables (PCAIV) [\citet{rao64}], and
particularly Canonical Correspondence Analysis and Redundancy Analysis,
which are primordial in ecological data analysis:
\begin{displaymath}
\begin{array}{c}
\xymatrix@L=8pt{
\fbox{$\mathbb{R}^p$} \ar[d]_{\mathbf{V}^{-1}_{11}} & \fbox{$\mathbb{R}^{{n}^*}$}
\ar[l]_{\mathbf{X}^T} \ar[r]^{{\mathbf{Y}}^T} & \fbox{$\mathbb{R}^q$} \ar
[d]^{\mathbf{D}_q\,.} \\
\fbox{$\mathbb{R}^{{p}^*}$} \ar[r]_{\mathbf{X}} & \fbox{$\mathbb{R}^n$} \ar
[u]_{\mathbf{D}_n} & \fbox{$\mathbb{R}^{{q}^*}$} \ar[l]^{\mathbf{Y}} \\}
\end{array}
\end{displaymath}

In instrumental variables methods, the Mahalanobis metric is used in
only one of the two spaces, most often the environmental variables
space~($\mathbb{R}^p$). This corresponds to the situation where one
wants to explain species distribution by linear combinations of
environmental variables, and this leads to the usual PCAIV triplet:
($\mathbf{Y}^T\mathbf{D}_n\mathbf{X}, (\mathbf
{X}^T\mathbf{D}_n\mathbf{X})^{-1}, \mathbf{D}_q$).

\subsection{Partial triadic analysis}

Partial Triadic Analysis [\citet{thioulouse87}] belongs to the STATIS
family of the $k$-table analysis methods [see, for example, the special
issue of the journal \textit{Comput. Statist. Data Anal.}
  \textbf{18}(1) (1994), or \citet{stanimirova04}]. The STATIS
family can be thought of as providing a PCA of a set of PCA's. In
ordinary PCA, the data table is summarized by a vector (principal
component), and in STATIS methods, the $k$-table is summarized by a
matrix. Partial Triadic Analysis is the most simple of these methods,
but it is also the most restrictive one. Its aim is to analyze a series
of $k$ tables having the same rows and the same columns. This means
that the same variables must be measured at the same sampling sites,
several times. Partial Triadic Analysis, like any STATIS-like method,
follows three steps: interstructure, compromise and intrastructure
(also called ``trajectories'').

The interstructure step provides the coefficients of a special linear
combination of the data tables, leading to an optimal summary called
the ``compromise.'' The second step computes the PCA of this linear
combination. The intrastructure step is a projection of the rows and
columns of each table of the series into the multidimensional space of
the compromise analysis.

The interstructure is based on the concepts of ``vector variance'' and
``vector covariance'' [\citet{escoufier73}]. It constructs a matrix of
scalar products between tables (the vector covariance matrix) that can
be written simply $\operatorname{Covv}(\mathbf{X}_k, \mathbf{X}_l)=\operatorname{Trace}(\mathbf
{X}^T_k\mathbf{D}_n\mathbf{X}_l\mathbf{D}_p)$, where $\mathbf{X}_k$
is the $k${th} table from the series. The eigenanalysis of this vector
covariance matrix gives a first eigenvector, and the components $\alpha
_k$ of this first eigenvector are used as weights to compute the compromise.

Alternatively, a ``vector correlation'' matrix can be used, that
rescales the tables: $\operatorname{Rv}(\mathbf{X}_k, \mathbf{X}_l) = \operatorname{Covv}(\mathbf
{X}_k, \mathbf{X}_l)/ \sqrt{\operatorname{Varv}(\mathbf{X}_k) \operatorname{Varv}(\mathbf
{X}_l)}$. $\operatorname{Varv}(\mathbf{X}_k)$ is the vector variance of table $k$:
$\operatorname{Varv}(\mathbf{X}_k) = \operatorname{Trace} (\mathbf{X}^T_k\mathbf{D}_n\mathbf
{X}_k\mathbf{D}_p)$. It is simply the usual variance of the vector
obtained by putting all the columns of table $\mathbf{X}_k$ one below
the other.

The compromise $\mathbf{X}_c$ is a linear combination of the initial
tables, weighted by the components of the first eigenvector of the
interstructure: $\mathbf{X}_c=\break\sum_k{\alpha_k\mathbf{X}_k}$. The
inertia of this compromise is maximized, and its main property is that
it maximizes the similarity with all the initial tables, as measured by
the sum of their squared dot product, $\sum_k\langle{\mathbf
{X}_c,\mathbf{X}_k} \rangle^2 $. When the tables are normed, the dot
product is the $\operatorname{Rv}$ coefficient.

The weight of each table is proportional to its inertia, so tables that
are different from the others will be downweighted. This property
ensures that the compromise will resemble all the tables of the
sequence ``as closely as possible'' in a least square sense. The
analysis of this compromise, for example, by a PCA, gives
two-dimensional representations (principal axes maps) that can be used to
interpret its structure.

The intrastructure is obtained by projecting the rows and columns of
each table of the series in the analysis of the compromise ({i.e.},
 the rows are projected on the principal axes, and the columns
are projected on the principal components). This step is done in the
same way as the projection of supplementary elements in a simple PCA
[see, for example, \citet{lebart84}, page~14]. Let $\mathbf{U}$ be
the matrix of the eigenvectors of the analysis of the compromise. The
scores of the rows of table $\mathbf{X}_k$
are $\mathbf{R}_k=\mathbf{X}_k\mathbf{D}_p\mathbf{U}$, and the
coordinates of its columns are $\mathbf{C}_k=\mathbf{X}^T_k\mathbf
{D}_n\mathbf{X}_c\mathbf{D}_p\mathbf{U}\bolds{\Lambda}^{-1/2}$,
where $\bolds{\Lambda}^{-1/2}$ is the diagonal matrix of the inverse
of the square root of the eigenvalues of the analysis of the compromise.

The advantage of Partial Triadic Analysis is that it highlights the
``stable structure'' in a sequence of tables. The compromise step
displays this stable structure (when it exists), and the intrastructure
step shows how each table moves away from~it.

\section{Data cube coupling methods}\label{sec3}

In this section we present three methods for analyzing a pair of data
cubes: BGCOIA, STATICO and COSTATIS. The principle of each method is
briefly explained, and the result obtained on the example data set is detailed.

\begin{figure}[t]

\includegraphics{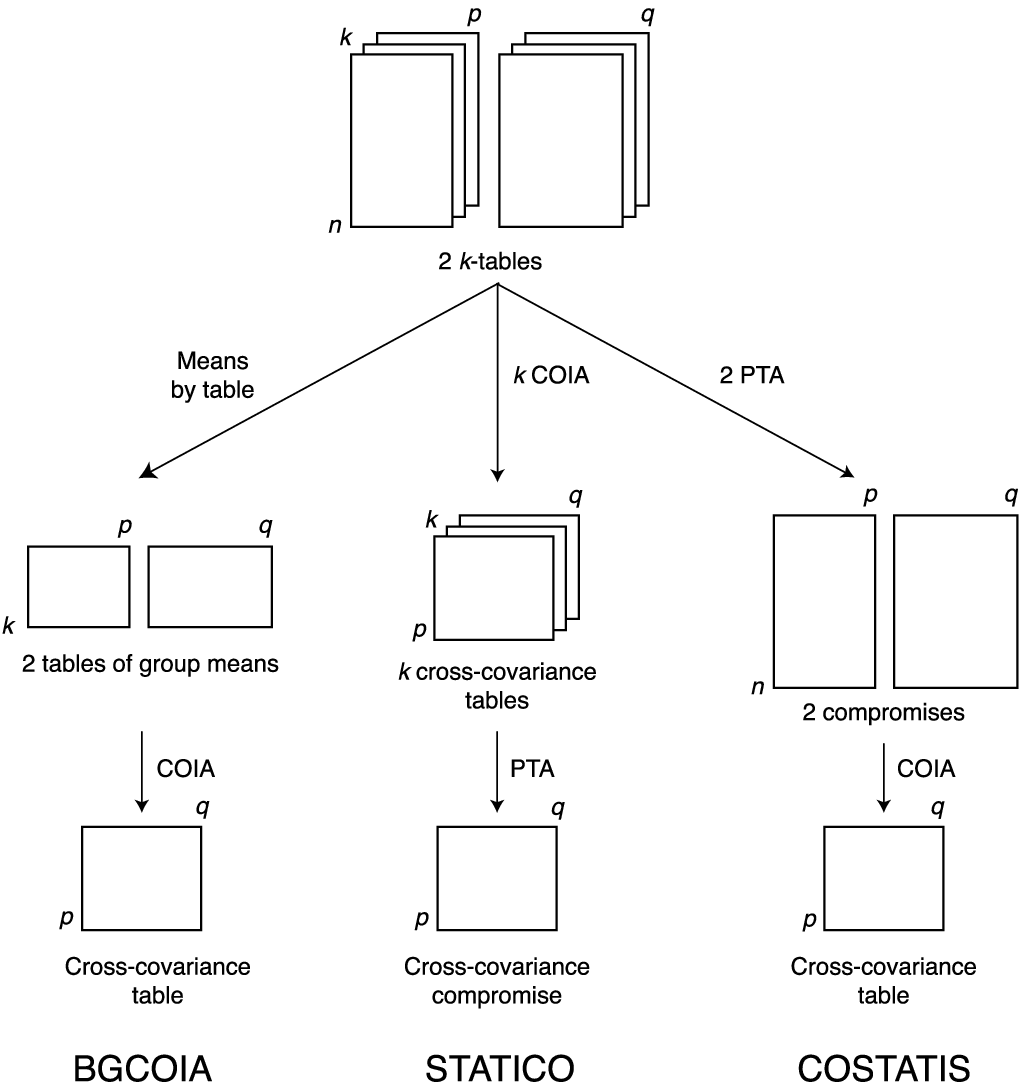}

\caption{Comparison of the three setups. BGCOIA is a between-group
co-inertia analysis, considering each table (date or site) of the
sequence as a group. STATICO is a Partial Triadic Analysis on the
series of cross product tables obtained by crossing the two tables of a
pair at each date. COSTATIS is a co-inertia analysis of the compromises
computed by the Partial Triadic Analysis of the two k-tables. In
BGCOIA, the mean of the variables in each table is computed and
arranged in two new tables. A~Co-Inertia Analysis is then done on this
couple of new tables. In STATICO, k cross-covariance tables are
computed from the two k-tables, resulting in a new k-table. A Partial
Triadic Analysis is then done on this new k-table. In COSTATIS, two
Partial Triadic Analyses are used to compute the compromises of the two
k-tables. A Co-Inertia Analysis is then used to analyze the
relationships between these two compromises.}\label{fig3}
\end{figure}

Figure~\ref{fig3} shows a comparison of the three approaches. BGCOIA is a~bet\-ween-group co-inertia analysis. It is therefore simply computed by
doing a Co-Inertia Analysis on the two tables of group means,
considering each table as a group [\citet{franquet95}]. In STATICO, we
first use Co-Inertia Analysis $k$ times to compute the sequence of $k$
cross-covariance tables, and then Partial Triadic Analysis to analyze
this new $k$-table. In COSTATIS, we first use two Partial Triadic
Analyses to compute the compromises of the two $k$-tables, and then
Co-Inertia Analysis to analyze the relationships between these two compromises.

\subsection{BGCOIA}

Let $g$ be the number of groups. The table of group means for
environmental variables is obtained by computing the means of each
variable within each group. This gives a new table, with $g$ rows and
$p$ columns. The same computations are done for the species data table,
leading to a~second new table with $g$ rows and $q$ columns. A simple
Co-Inertia Analysis is performed on these two new tables. The rows of
the initial tables can be projected into this analysis to help
interpret the results [\citet{lebart84}].

The main advantage of this method is its simplicity, from both
theoretical and practical points of view. The two data cubes are
reduced to two tables by taking the means of the columns of each
elementary table of the cubes. Co-Inertia Analysis is then applied to
the two resulting tables.

The fact that it is a Between-Group Analysis can be used to give more
importance to the spatial or the temporal effect in a space--time
experimental design. Tables can correspond to dates or to sampling
sites and, depending on the importance that should be given to space or
time processes, one or the other of the two possibilities can be used.
On the example data set used here, \citet{franquet95} considered that
one table corresponds to one sampling site, so we use the same setup.

Note that the symmetric method, Within-Group Co-Inertia Analysis
(WGCOIA) based on Co-Inertia Analysis and Within-Group Analysis [\citet
{franquet94}] can also be used to analyze one effect after removing the
other, which results in four possible setups and four different
analyses (between dates, between sites, within dates and within sites).

\begin{figure}

\includegraphics{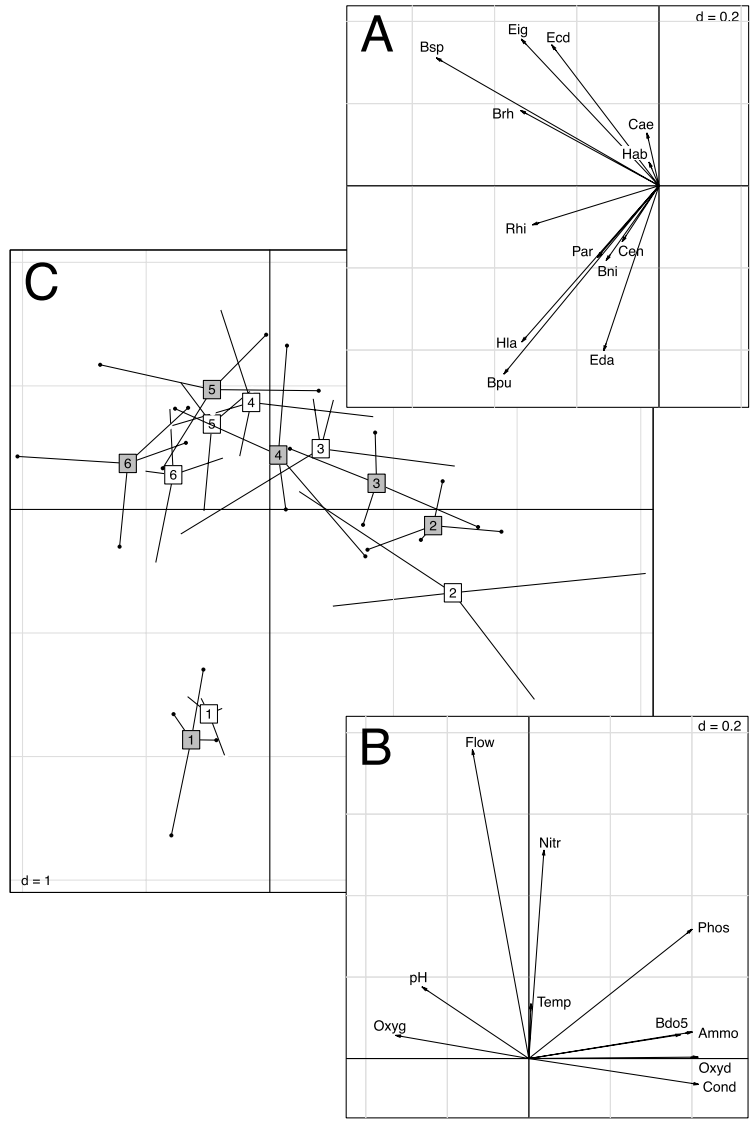}

\caption{First two principal axes maps of the between-group co-inertia
analysis (see text for details). The eigenvalues corresponding to these
two axes are equal to 70.2 and 4.45. The scale is given by the value
(d) in the upper right or lower left corner of each plot; it
corresponds to the size of the background grid. \textup{(A)}~Map of the rows of
the cross product table (Ephemeroptera species). \textup{(B)}~Map of the columns
(environmental variables). \textup{(C)}~Map of the sites. The 48 points in
\textup{(C)} correspond to the 6 sites (labeled 1--6), sampled 4 times, with one
set of points for the environmental variables table sequence (open
circles, white background site labels) and one set of points for the
Ephemeroptera species table sequence (black circles, gray background
site labels).}\label{fig4}
\end{figure}

Figure~\ref{fig4} shows the results of the between-group co-inertia analysis; it
corresponds to Figure~4 of \citet{franquet95}. Figure~\ref{fig4}A is the
principal axes map of the rows of the cross product table
(Ephemeroptera species), Figure~\ref{fig4}B is the principal axes map of the
columns (environmental variables), and Figure~\ref{fig4}C is the principal axes
map of the sites. Figures~\ref{fig4}A and~\ref{fig4}B are obtained directly with the row
scores and column loadings of the cross product table, while Figure~\ref{fig4}C
is obtained by projecting the rows of the two sequences of tables as
supplementary elements into the co-inertia analysis space.

The 48 points on Figure~\ref{fig4}C correspond to the 6 sites, sampled 4 times,
and there is one set of points for the environmental variables table
sequence (open circles) and one set of points for the Ephemeroptera
species table sequence (black circles). The columns of these two
sequences of tables (environmental variables at each date and
Ephemeroptera species at each date) could also be projected into the
co-inertia analysis, but this has not been done here.

The four points corresponding to the four sampling dates for each site
are grouped to form a star, and the barycenter of these four points is
labeled with the number of the site (white background label for the
environmental variables table sequence, gray background label for the
Ephemeroptera species table sequence).

The interpretation of Figure~\ref{fig4}C is simple. The first axis (Figure~\ref{fig4}B,
horizontal) is a pollution gradient from left (unpolluted situation:
high concentration of oxygen and high pH) to right (highly polluted
situation: high concentrations of amonium and phosphates, high
conductivity and oxydability, high biological oxygen demand). The
second axis (Figure~\ref{fig4}B, vertical) is an upstream-downstream physical gradient:
discharge (``Flow'') increases downstream (upward on the figure). Most
Ephemeroptera species are more abundant in unpolluted situations (Figure~\ref{fig4}A,
horizontal), and some species are characteristic of the lower part of
the stream (Bsp, Eig, Ecd), while others (Bpu, Hla, Eda) are
characteristic of the upper part, or of site 6 (Figure~\ref{fig4}A, vertical).

Figure~\ref{fig4}C shows that the spatial component of the phenomenon is more
important than the temporal aspect. The first axis (Figure~\ref{fig4}C, horizontal)
opposes unpolluted sites (site 1, upstream the Autrans summer mountain
resort, and site 6) to highly polluted sites (site 2, just downstream
Autrans, and receiving the ouputs of the sewer system). Sites 3, 4 and
5 are less and less polluted. This corresponds to the natural
restoration process: the pollution is gradually resorbed along the
stream. Figure~\ref{fig4}C also shows that the biological processes are very
linked to the physico-chemical variations of water quality along the
stream: the two ordinations of sites are very similar.

\subsection{STATICO}
The STATICO method is based on the Partial Triadic Analysis of a
sequence of cross product tables (see Figure~\ref{fig3}). Starting from the
sequence of paired ecological tables, each cross product table is
computed using the pair of tables at each date. All the tables of the
sequence do not need to have the same number of rows, but they need to
have the same number of columns across dates. This means that the
number of sampling sites can vary from one date to another, but the
number of environmental variables ($p$) must be the same for all the
dates, and the number of species~($q$) must also be the same for all
dates. Therefore, all the cross product tables have the same number of
rows ($p$) and columns ($q$). They contain the covariances between the
columns of the two tables.

Let ($\mathbf{X}_k, \mathbf{D}_p, \mathbf{D}_{n_k}$) and ($\mathbf
{Y}_k, \mathbf{D}_q, \mathbf{D}_{n_k}$) be the pair of triplets at
date $k$. $\mathbf{X}_k$ is the table of environmental variables
measured at date $k$, and $\mathbf{Y}_k$ is the table of species
observed at the same date. $\mathbf{D}_p$ and $\mathbf{D}_q$ are the
same for all the dates and $\mathbf{D}_{n_k}=\operatorname{Diag}(\frac{1}{n_k})$ is
the same for both tables. Let $\mathbf{Z}_k$ be the $k${th} cross
product table: $\mathbf{Z}_k=\mathbf{Y}_k^T\mathbf{D}_{n_k}\mathbf
{X}_k$. The Co-Inertia Analysis triplet at date $k$ is ($\mathbf{Z}_k,
\mathbf{D}_p, \mathbf{D}_q$) and the STATICO method is the Partial
Triadic Analysis of the $k$-table made by this series of cross product tables.

The interstructure step gives optimal weights $\alpha_k$ such that the
inertia of the triplet $(\sum_{k} \alpha_k\mathbf{Z}_k, \mathbf
{D}_p, \mathbf{D}_q)$ is maximum with the constraint $\sum_{k} \alpha_k^2=1$.

The compromise of the STATICO method $(\mathbf{Z})$ is a weighted mean
of the cross product tables using weights $\alpha_k$: $\mathbf
{Z}=\sum_{k} \alpha_k\mathbf{Z}_k$ [\citet{simier99}]. The analysis
(PCA) of this compromise gives a graphical display of the environmental
variables (rows of $\mathbf{Z}$) and of the species (columns of
$\mathbf{Z}$).

Finally, the intrastructure step projects the rows and columns of each
table of the sequence in the analysis of the compromise, with usual
supplementary element projection technique [\citet{lebart84}]. This
gives a display of the environmental variables at each date, of the
species at each date, and two displays of the sampling sites at each
date (one from the point of view of environmental variables and one
from the point of view of species).

\begin{figure}[t]

\includegraphics{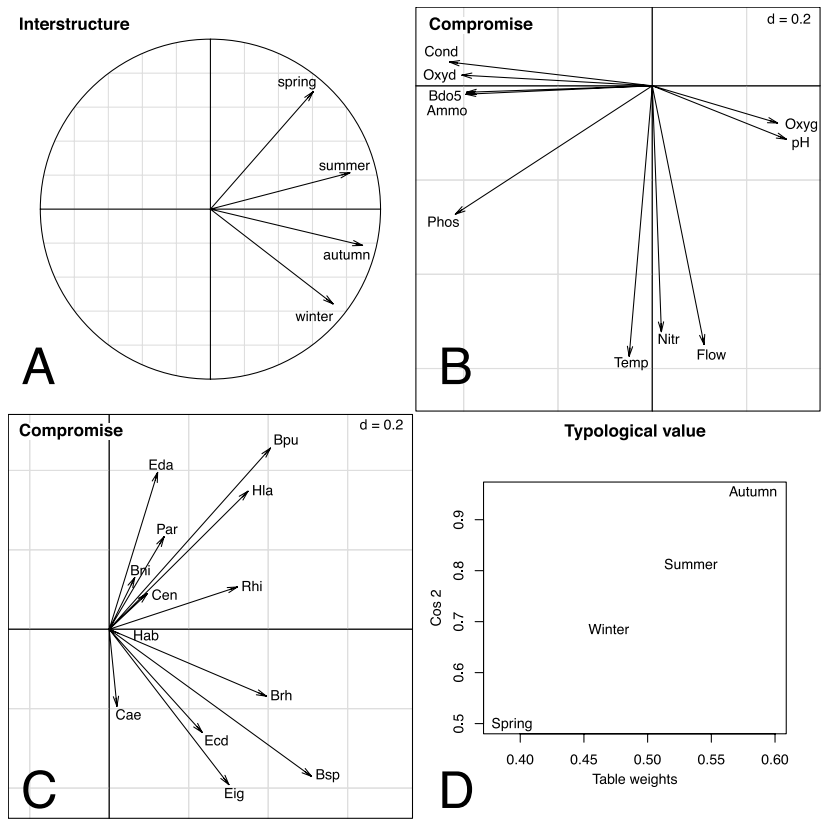}

\caption{Results of the STATICO method. This is a compound graph,
automatically drawn by the ``plot'' function of the ade4 package for a
partial triadic analysis. The eigenvalues corresponding to the two axes
of the compromise analysis are equal to 593.6 and 45.3. The scale is
given by the value (d) in the upper right corner of the compromise
plots; it corresponds to the size of the background grid. The four
plots are as follows: \textup{(A)}  The interstructure plot, showing the four
seasons, and the importance of the corresponding tables in the
definition of the compromise (coordinate of the points on the first
axis). \textup{(B)} Compromise analysis principal axes map (environmental
variables). \textup{(C)} Compromise analysis principal axes map (Ephemeroptera
species). \textup{(D)} Typological values of the four tables (square cosines vs.
table weights).}\label{fig5}
\end{figure}

\begin{figure}[t]

\includegraphics{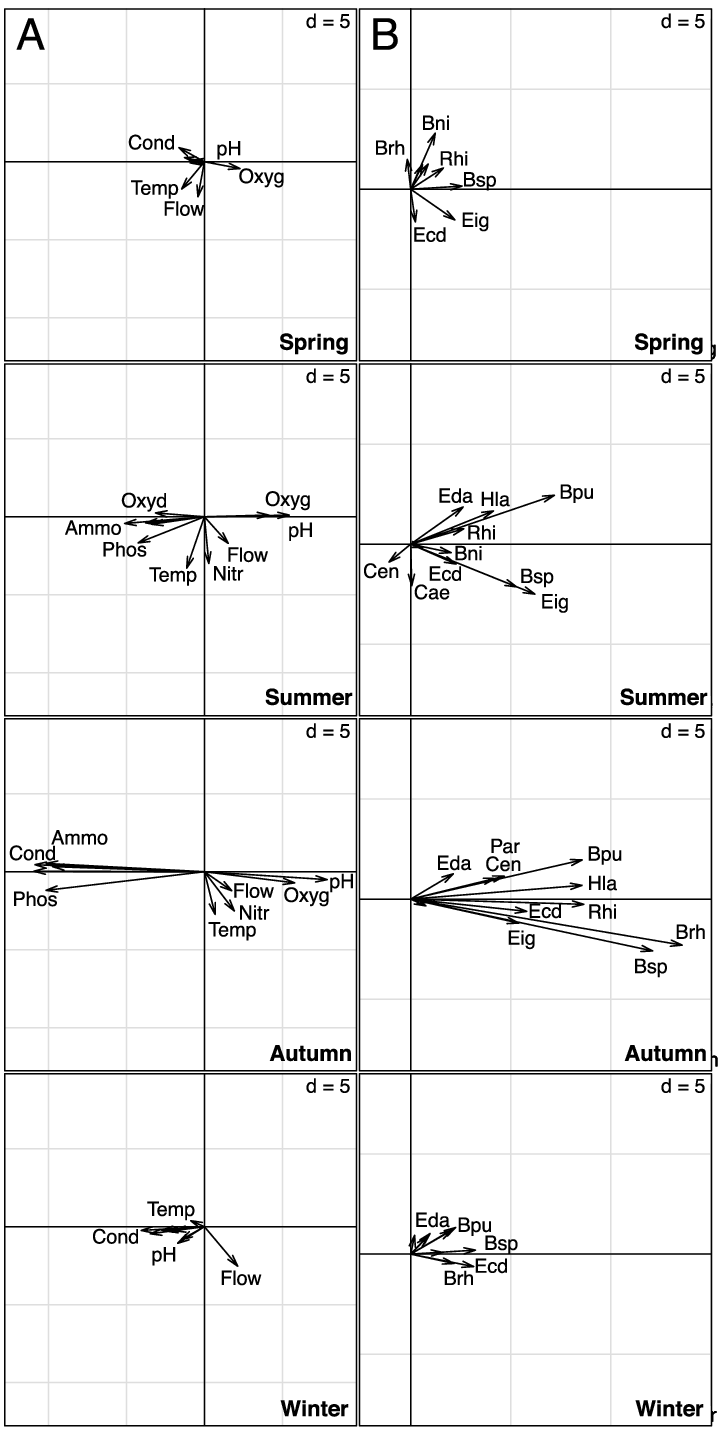}%
\vspace*{-3pt}
\caption{Results of the STATICO method: variable intrastructure step.
The scale is given by the value (d) in the upper right corner of each
plot; it corresponds to the size of the background grid. The four plots
on the left \textup{(A)} show the environmental variables at each date (Spring,
Summer, Autumn, Winter), and the four plots on the right \textup{(B)} show the
Ephemeroptera species at the same dates.}\label{fig6}
\vspace*{-5pt}
\end{figure}

\begin{figure}[t]

\includegraphics{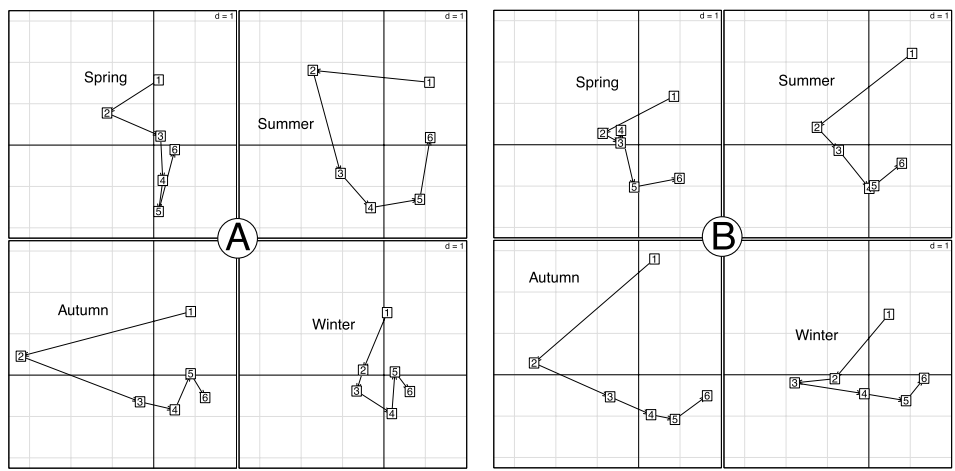}

\caption{Results of the STATICO method: site intrastructure step
(``trajectories''). The scale is given by the value (d) in the upper
right corner of each plot; it corresponds to the size of the background
grid. \textup{(A)} trajectories of sites for environmental variables.
\textup{(B)}~trajectories of sites for Ephemeroptera species. These plots show the
distortions of the upstream-downstream gradients across seasons, as
Ephemeroptera species react to pollution increase (maximum reached in
Autumn) or decrease (minimum in Spring).}\label{fig7}
\end{figure}

The results of the STATICO method are presented in Figures~\ref{fig5}--\ref{fig7}.
Figure~\ref{fig5} is a compound graph that sums up the first two steps of the
STATICO method: interstructure and compromise. Figure~\ref{fig6} shows the
intrastructure step for environmental variables and Ephemeroptera
species ({i.e.}, the projection of the columns of the tables of
the two sequences as supplementary elements in the compromise
analysis), and Figure~\ref{fig7} is the intrastructure step for the sites
({i.e.}, the projection of the rows of the tables of the two sequences).

The STATICO method is a Partial Triadic Analysis on the sequence of
cross product tables, so the compromise is also a cross product table,
with the 13 Ephemeroptera species in rows and the 10 environmental
variables in columns. Sites have disappeared from this table, but they
can be projected as supplementary elements to help interpret the
results of the analysis.

The interstructure plot (Figure~\ref{fig5}A) shows that Autumn and Summer are
the two most important seasons for defining the compromise, while
Winter and Spring are slightly less important.

The compromise plots (Figure~\ref{fig5}B and~C) are very similar to the BGCOIA
plots (Figure~\ref{fig5}A and~B). They show that the first axis (horizontal)
is also a~pollution gradient: clean water on the right, and pollution
on the left. The second axis (vertical) is also an upstream--downstream
physical gradient: discharge (``Flow'') and temperature (``Temp'')
increase downstream (downward on the figure). Nitrates (``Nitr'') also
increase along the whole stream instead of having a maximum at site 2
like other pollution variables, and this is why they are located here.
The sensitivity of all Ephemeroptera species to pollution and the
specificity of some species (Bpu, Hla, Eda upstream and Bsp, Eig, Ecd
downstream) are also found again.\vadjust{\goodbreak}

The ``typological value'' plot (Figure~\ref{fig5}D) shows that Autumn has the
highest influence in the construction of the compromise, while Spring
has the lowest.

Figure~\ref{fig6} shows the intrastructure step for environmental variables
(Figure~\ref{fig6}A) and Ephemeroptera species (Figure~\ref{fig6}B). It is drawn using
the projection of the columns of the two sequences of tables as
supplementary elements in the compromise analysis.

Autumn is clearly the season where the structures are the strongest
(arrows are much longer at this date), both for environmental variables
and for Ephemeroptera species. Conversely, Spring is the season where
the structures are the weakest (arrows are all very short). This
confirms the interpretations made on Figure~\ref{fig5}. However, although the
structures may vary in intensity, they are preserved across dates: the
first axis is always a pollution gradient, and the second one is always
an upstream--downstream opposition.

Figure~\ref{fig7} shows the intrastructure step for the sites. It is drawn using
the projection of the rows of the two sequences of tables as
supplementary elements in the compromise analysis. It is very similar
to Figure~\ref{fig4}C, but it is split according to seasons instead of sites.
The environmental variable plot (Figure~\ref{fig5}A) and the Ephemeroptera
species plot (Figure~\ref{fig7}B) are placed side by side. This presentation
insists on the comparison between the four seasons, showing mainly the
distortions of the upstream--downstream gradient across seasons.

The differences among the seasons are shown clearly in Figure~\ref{fig7}. In
Spring, sites are lined up on the upstream-downstream gradient and only
site 2 moves slightly to the left. In Summer, pollution is highest at
site 2, and restoration occurs along sites 3, 4 and 5. In Autumn,
pollution is maximum because stream flow is at its minimum (pollutant
concentrations are maximum). In Winter, pollution has almost
disappeared, because Autrans is a Summer mountain resort, but the
upstream--downstream gradient is still disturbed.

Figure~\ref{fig7}B shows the same structures, because the pollution has a
negative impact on Ephemeroptera species abundance (horizontal axis)
and because of the upstream--downstream preferences of particular
species (vertical axis).

\subsection{COSTATIS}

COSTATIS is a new method that is also based on $k$-table methods and on
co-inertia, but it benefits from the advantages of both STATICO and
BGCOIA. Indeed, it has the same optimality properties of k-table
analyses as STATICO ({i.e.}, the maximizing properties of the
compromise), but it has the simplicity of BGCOIA.

COSTATIS is simply a co-inertia analysis of the compromises of two
$k$-table analyses (see Figure~\ref{fig3}). The first step of COSTATIS consists in
performing two Partial Triadic Analyses: one on the environmental
variables $k$-table, and one on the species $k$-table. The second step is
simply a co-inertia analysis of the compromises of these two Partial
Triadic Analyses. This means that the number of tables does not have to
be the same for the two series of tables, but that the number of
species, of environmental variables, and of sampling sites must be the
same for all the tables.

$\mathbf{X}_c=\sum_k{\alpha_k\mathbf{X}_k}$ is the ($n \times p$)
compromise of the Partial Triadic Analysis of environmental variables,
and $\mathbf{Y}_c=\sum_k{\beta_k\mathbf{Y}_k}$ is the ($n \times q$)
compromise of the Partial Triadic Analysis of species data. These
compromises are weighted means of the tables of the original sequences,
with weights equal to the components of the first eigenvector of the
interstructure of the two Partial Triadic Analyses. The inertias of the
triplets $(\mathbf{X}_c, \mathbf{D}_p, \mathbf{D}_n)$ and $(\mathbf
{Y}_c, \mathbf{D}_q, \mathbf{D}_n)$ are maximum under the constraints
$\sum_{k} \alpha_k^2=1$ and $\sum_{k} \beta_k^2=1$.

The Co-Inertia Analysis of these two compromises decomposes the total
co-inertia:
\[
\mathit{CoI}_{\mathbf{X}_c\mathbf{Y}_c}=\operatorname{trace}(\mathbf{X}_c\mathbf
{D}_p\mathbf{X}_c^T\mathbf{D}_n\mathbf{Y}_c\mathbf{D}_q\mathbf
{Y}_c^T\mathbf{D}_n),
\]
and maximizes the co-inertia between species and environmental variable
scores. An additional step can be implemented, like in the STATICO
method: it is possible to project the rows and columns of all the
tables of the two sequences as supplementary elements into the
multidimensional space of this Co-Inertia Analysis.

Each compromise represents the ``stable structure'' of the
corresponding sequence: $\mathbf{X}_c$ is the stable structure of the
environmental tables sequence, and $\mathbf{Y}_c$ is the stable
structure of the species tables sequence. COSTATIS brings to light the
relationships between these two stable structures, and it discards the
conflicting variations between the whole sequences. It is therefore
very easy to interpret (like a standard Co-Inertia Analysis), yet it
retains the optimality properties of the compromises of the two Partial
Triadic Analyses.

\begin{figure}

\includegraphics{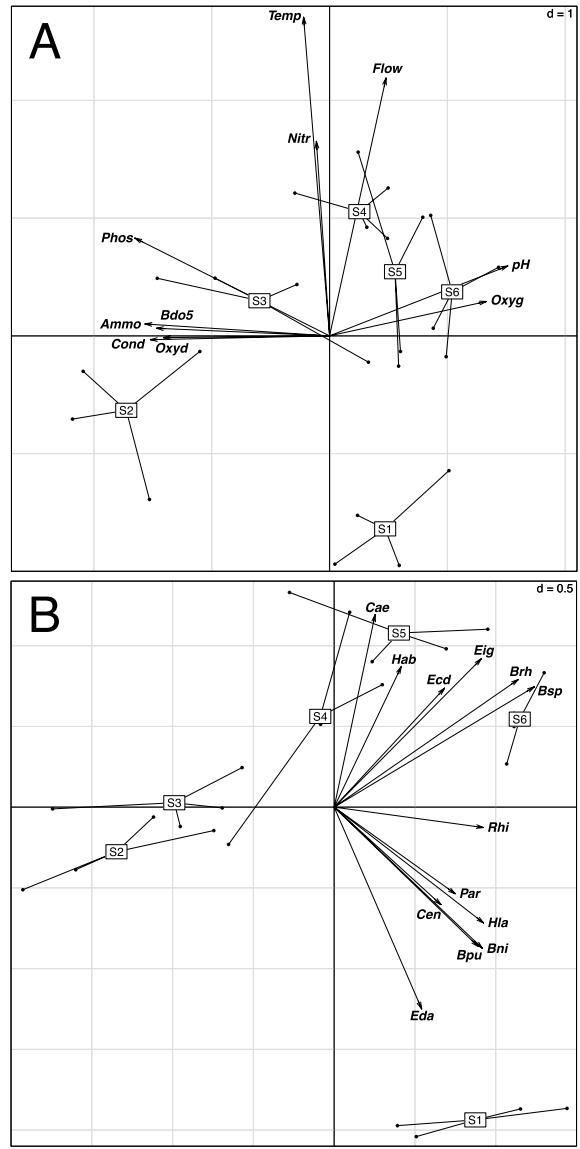}

\caption{Results of the COSTATIS method (first two axes maps). The
eigenvalues corresponding to the two axes are equal to 34.52 and 6.695.
The scale is given by the value (d)~in the upper right corner of each
plot; it corresponds to the size of the background grid. \textup{(A)}~First
biplot of the co-inertia analysis between the compromises of the two
Partial Triadic Analyses, with sites and variables superimposed. \textup{(B)}
Second biplot, with sites and species superimposed. These two biplots
could be superimposed.}\label{fig8}
\end{figure}

COSTATIS results are presented in Figure~\ref{fig8}. COSTATIS is a co-inertia
analysis, and it is therefore possible to use a permutation test to
assess the statistical significance of the relationships between the
two tables, just like in a usual Co-Inertia Analysis. The result of
this permutation test on the M\'eaudret data set gave a $p$-value of 1\%.

The co-inertia analysis is done on the compromises of two $k$-table
analyses. Here, we used two Partial Triadic Analyses, but the results
of these two analyses are not presented. We show only the plots of the
co-inertia analysis, under the form of two biplots: one for
environmental variables (Figure~\ref{fig8}A), and one for Ephemeroptera species
(Figure~\ref{fig8}B). These two biplots could, in fact, be superimposed on the
same figure, but the outcome was cluttered.

Presenting the results in this way underlines the fact that COSTATIS is
looking for the relationships (co-structure) between the stable
structures extracted from two series of tables. Figure~\ref{fig8}A shows the
results for the environmental variables. The same structure as the one
detected by STATICO and BGCOIA is observed. Axis 1 is the pollution
gradient (pollution on the left) and axis 2 is the upstream--downstream
opposition (downstream is upward).

The four dates for each site are projected on this plot and, like in
the BGCOIA plot (Figure~\ref{fig4}C), the four points corresponding to the four
sampling dates of each site are grouped to form a star. The gravity
center of these four points is labeled with the number of the site. The
four points of site 2 are on the left, as pollution is higher in this
site for the four dates (except for site 3 in Winter). Pollution
decreases downstream along sites 3, 4 and 5, and is the lowest at site 6.

The second biplot is presented in Figure~\ref{fig8}B. It shows the Ephemeroptera
species, with the same opposition between upstream and downstream
characteristic species. In the same way as in Figure~\ref{fig8}A, the four dates
for each site are projected on the plot and the corresponding four
points are grouped to form a star. The gravity center of these four
points is labeled with the number of the site. The position of sites
corresponds to the abundance of the species in these sites: sites 2 and
3 have the lowest number of Ephemeroptera, so they are far on the left.
Site 1 has the highest number of species ``Eda,'' and sites 5 and 6
have the highest number of species ``Bsp,'' ``Brh'' and
``Eig.''\looseness=-1

The first axis common to these two biplots ({i.e.}, the first
COSTATIS axis) maximizes the covariance between the coordinates of the
``compromise variables'' and the ``compromise species.'' The result is
that it displays the relationships\vadjust{\goodbreak} between the stable structures
extracted from two data sets. On this example, this relationship is the
fact that the pollution gradient affects the abundance of Ephemeroptera
species. The second axis represents the upstream-downstream opposition,
and the relationships between ecological preferences of Ephemeroptera
species and physical variables or stream morphology.

\section{Discussion}\label{sec4}

The three methods presented here uncover the same features in the
example data set. This is a small data set, but with strong structure,
and strong structures often are clear with any method. However, the
three methods used to analyze even a data set with clear structure can
have advantages and drawbacks. The advantages of these methods can be
summarized as follows:

\begin{description}
\item[BGCOIA:] It is the most straightforward method. It is simple to
apply and outputs are easy to interpret. It can be used to favor one
point of view (for example, space {vs.} time), by choosing the
factor of Between-Group Analysis. It can also be used in conjunction
with WGCOIA to study an effect (time) after removing the other (space).

\item[STATICO:] The main advantage of this method is the optimality of
the compromise (maximization of the similarity with all the initial
tables). It gives a compromise of co-structures, which means that it
displays the stable component of species--environment relationship
variations. It benefits from the three-steps computation scheme of
STATIS-like methods (interstructure, compromise, intrastructure), and
graphical outputs can be very detailed.

\item[COSTATIS:] This method benefits from the advantages of the two
others: optimality of the Partial Triadic Analysis compromises, ease of
use, simplicity of co-inertia analysis graphical outputs. COSTATIS is
the co-inertia analysis of two compromises, so it looks for the
relationships between two stable structures. This is different from the
STATICO point of view (co-structure of two compromises {vs.}
compromise of a series of co-structures).
\end{description}

The three methods can also be compared from the perspective of the
possible objectives of a data cube coupling strategy. The first
objective is to find a ``consensus'' in the relationships between
species and environment. This consensus should be independent from the
repetitions (time or space), and the three methods achieve this in
different ways.

In COSTATIS, a consensus is extracted first, separately and
independent\-ly for environmental variables and species. The
relationships between~these two summaries are then investigated by a
co-inertia analysis. In STATICO, species--environment relationships are
first analyzed at each date, and a~stab\-le summary of these
relationships is then computed.

If species--environment relationships are weak, or present only at some
dates, they may disappear after the first step of COSTATIS (the two
separate Partial Triadic Analyses) and the final co-inertia analysis
permutation test may be nonsignificant. Conversely, if
species--environment relationships are very strong, chronological
structures may disappear in STATICO. COSTATIS should therefore be
preferred when species--environment relationships are strong and
chronological structures are not of primary importance.

Another objective of data cube coupling strategy, complementary to the
first one, can be the search for a description of the evolution of
species--environment relationships (like seasonal variations or long
term changes), rather than a description of the stable part of these
relationships. In this case, STATICO may be more appropriate than
COSTATIS, as it computes a~consensus of species--environment
relationships at each date, and only after builds a time consensus.

BGCOIA is slightly different, because it makes easy a choice in the
initial analysis between a spatial or a chronological setup. It should
be used only when there are good reasons to give the priority to space
or to time. But on the other hand, WGCOIA can be used after the BGCOIA,
to remove the primary effect (space or a time) chosen in BGCOIA.

Great care must be taken in the choice of the factor defining the
groups for the BGCOIA method. \citet{franquet95} explain why they chose
a between-site (as opposed to between-season) co-inertia analysis on
the example data set. In Hydrobiology, seasonal variations are mostly
linked to water temperature, and the corresponding between-season
structures are trivial (Summer--Winter opposition). But in other
situations, a between-date analysis could be an interesting strategy.
This choice of the grouping factor, combined with the possibility to
use WGCOIA after a BGCOIA gives four different analyses (between-site,
between-date, within-sites and within-date) that can be used to
explore complex data sets.

For $k$-table methods, the way of organizing the $k$-table in a series of
\textit{k} tables is also important. For a three-way array
(sites${}\times{}$variables${}\times{}$dates), there are three ways to cut the data cube into a
series of tables. However, only two are really interesting. Indeed, the
option ``one table $=$ one variable'' is not coherent with the aim of the
analysis: a compromise between physico-chemical variables would have no
meaning. So we have to choose between ``one table $=$ one date'' (as done
in the present paper) or ``one table $=$ one site.'' This choice is
dictated by the objectives of the study and also by the fact that the
method will try to compute a compromise as a linear combination of the
tables. This means that this compromise should be meaningful.

A third point of comparison between data cube coupling methods is the
numerical constraints put on the parameters of the $k$-table,
{that is}, the number of species, of environmental variables, of
sampling sites, and of dates. From this point of view, BGCOIA, STATICO
and COSTATIS share the same constraints on the species and
environmental variables, which should always be identical: same species
and same environmental variables for all dates.\vadjust{\goodbreak}

But the constraints are different for dates and sampling sites. In
COSTATIS, the two series of tables can have different numbers of dates
(and even different dates), while the sampling sites must be the same
for all the tables at all dates. In STATICO, the two series of tables
must have the same dates, but sites can differ among dates (although
they must be equal for the two tables of a pair). Constraints from the
experimental design can therefore influence the choice of the method.

Moreover, the constraints on species and environmental variables come
from the choice of the $k$-table analysis in COSTATIS and STATICO (a
Partial Triadic Analysis). Extensions of these methods can be imagined,
that would use another variant of STATIS-like analyses instead of
Partial Triadic Analysis. For example, using STATIS on operators (the
classical ACT method) [\citet{lavit94}] would lead to a ``site
COSTATIS'' method allowing the use of both different species and
different environmental variables. We could also define a ``species
STATICO'' allowing the use of different environmental variables among
dates, and a ``variable STATICO'' allowing the use of different species
among dates.

This possibility of having varying species, environmental variables,
dates, and sampling sites makes the use of the three methods much more
flexible, but only COSTATIS allows the use of different species
{and} different environmental variables. However, it should be used
with care, as this flexibility might be obtained at the expense of
loosing some of the structures in the data set.

A drawback common to all these methods is the relative complexity of
exploratory multivariate data analysis. In this area, the ``ade4''
package for the R environment [\citet{rcoreteam}] tries to make things
easier. Simple function syntax and structured objects have been
privileged. In addition, a graphical user interface is available in the
``ade4TkGUI'' package [\url{http://pbil.univ-lyon1.fr/ade4TkGUI/},
\citet{ThioulouseDray2007}], and $k$-table methods will be implemented in
this interface. Moreover, all the computations and graphical displays
in this article can be redone interactively online, thanks to this
reproducibility page: \href{http://pbil.univ-lyon1.fr/SAOASOPET/}{http://pbil.univ-lyon1.}
\href{http://pbil.univ-lyon1.fr/SAOASOPET/}{fr/SAOASOPET/}.

The availability of methods able to analyze data sets with a complex
organization, like pairs of data cubes, is important because it allows
to take into account this organization and to analyze the data sets
globally. There are alternatives to these methods, like analyzing
stacked tables, or performing several separate analyses, like time
series analysis for each variable, or functional data analysis on each
table. But the analysis is facilitated by taking into account the data
structure as dictated by the experimental design. Exploring
species--environment relationships is not an easy task, and
adding
spatial and temporal influences makes it even more difficult, but this
is a~necessary step toward the comprehension of ecosystem functioning.

\section*{Acknowledgments}
The author would like to thank the Editor, the Associate Editor,
Thibaut Jombart, and an anonymous referee for many useful comments and
suggestions that helped improve the first version of this
paper.\looseness=-1


\printaddresses

\end{document}